\documentclass[journal=jpcl,manuscript=letter]{achemso}

\usepackage[version=3]{mhchem} 
\usepackage{upgreek}
\usepackage{threeparttable}
\usepackage{booktabs}
\usepackage{multirow}
\usepackage{hyperref}
\usepackage{color,soul}

\author{Lanhai He}
\affiliation{Center for Free-Electron Laser Science, Deutsches Elektronen-Synchrotron DESY, Notkestraße~85, 22607 Hamburg, Germany}
\alsoaffiliation{Institute of Atomic and Molecular Physics, Jilin University, 130012 Changchun, China}
\altaffiliation{These authors contributed equally.}
\author{Luk\'{a}\v{s}~Toman\'{i}k}
\affiliation{Department of Physical Chemistry, University of Chemistry and Technology, Technick\'{a} 5, 16628 Prague, Czech Republic}
\altaffiliation{These authors contributed equally.}
\author{Sebastian Malerz}
\affiliation{Molecular Physics, Fritz-Haber-Institut der Max-Planck-Gesellschaft, Faradayweg 4-6, 14195 Berlin, Germany}
\author{Florian Trinter}
\affiliation{Molecular Physics, Fritz-Haber-Institut der Max-Planck-Gesellschaft, Faradayweg 4-6, 14195 Berlin, Germany}
\alsoaffiliation{Institut für Kernphysik, Goethe-Universität Frankfurt, Max-von-Laue-Straße 1, 60438 Frankfurt am Main, Germany}
\author{Sebastian Trippel}
\affiliation{Center for Free-Electron Laser Science, Deutsches Elektronen-Synchrotron DESY, Notkestraße~85, 22607 Hamburg, Germany}
\alsoaffiliation{Center for Ultrafast Imaging, Universität Hamburg, Luruper Chaussee 149, 22761~Hamburg, Germany}
\author{Michal Belina}
\affiliation{Department of Physical Chemistry, University of Chemistry and Technology, Technick\'{a} 5, 16628 Prague, Czech Republic}
\author{Petr Slav\'{i}\v{c}ek}
\affiliation{Department of Physical Chemistry, University of Chemistry and Technology, Technick\'{a} 5, 16628 Prague, Czech Republic}
\email{Petr.Slavicek@vscht.cz}
\author{Bernd Winter}
\affiliation{Molecular Physics, Fritz-Haber-Institut der Max-Planck-Gesellschaft, Faradayweg 4-6, 14195 Berlin, Germany}
\email{winter@fhi-berlin.mpg.de}
\author{Jochen Küpper}
\affiliation{Center for Free-Electron Laser Science, Deutsches Elektronen-Synchrotron DESY, Notkestraße~85, 22607 Hamburg, Germany}
\alsoaffiliation{Center for Ultrafast Imaging, Universität Hamburg, Luruper Chaussee 149, 22761~Hamburg, Germany}
\alsoaffiliation{Department of Physics, Universität Hamburg, Luruper Chaussee 149, 22761 Hamburg, Germany}
\email{jochen.kuepper@cfel.de}

\title[An \textsf{achemso} demo]
  {Specific \textit{versus} Non-Specific Solvent Interactions of a Biomolecule in Water}

\begin{document}




\newpage
\begin{abstract}
   %
   \noindent%
   Solvent interactions, particularly hydration, are vital in chemical and biochemical systems.
   Model systems unveil microscopic details of such interactions. We uncover a specific
   hydrogen-bonding motif of the biomolecular building block indole (C$_8$H$_7$N), tryptophan's
   chromophore, in water: a strong localized $\text{N-H}\cdots\text{OH}_2$ hydrogen bond, alongside
   unstructured solvent interactions. This insight is revealed from a combined experimental and
   theoretical analysis of indole's electronic structure in aqueous solution. We have recorded the
   complete X-ray photoemission and Auger spectrum of aqueous-phase indole, quantitatively
   explaining all peaks through \emph{ab initio} modeling. The efficient and accurate technique for
   modeling valence and core photoemission spectra involves the maximum-overlap method and the
   non-equilibrium polarizable-continuum model. A two-hole electron-population analysis
   quantitatively describes the Auger spectra. Core-electron binding energies for nitrogen and
   carbon highlight the specific interaction with a hydrogen-bonded water molecule at the N-H group
   and otherwise nonspecific solvent interactions.
\end{abstract}

\begin{figure}
\centering
  \includegraphics[width=2in]{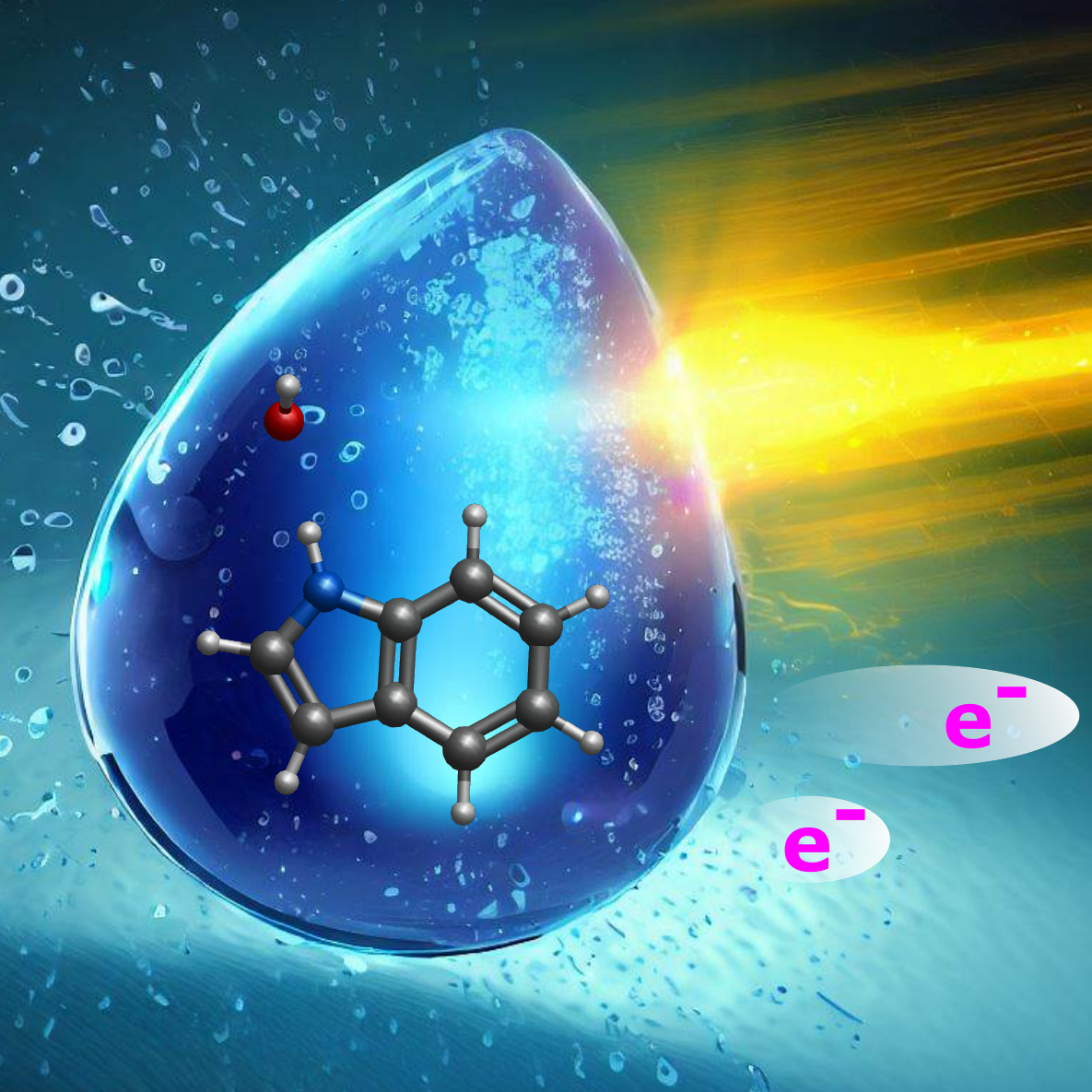}
\end{figure}


\noindent Indole (C$_8$H$_7$N) is a ubiquitous component of peptides and proteins as it is the side chain chromophore
of the essential amino acid tryptophan. Indole has various signaling
functions~\cite{Sobolewski:PCCP4:1093,Plekan:MP106:1143,Chrostowska:JACS136:11813}, 
and most relevant for the present work, it starts to absorb strongly below 300~nm,
with the main bands around 287 and 217~nm, both corresponding to $\pi\rightarrow\pi^\ast$
transitions.~\cite{Vivian:BiophysJ80:2093,Sarkisyan:SR2:608}. The influence of solvation on indole's electronic structure has long been
debated~\cite{Short:JCP108:10189}, in particular with respect to its two lowest-energy $^1L_a$ and
$^1L_b$ electronically excited singlet states~\cite{Callis:JCP95:4230,Brand:PCCP12:4968:2010,
   Kuepper:PCCP12:4980}. Microsolvation experiments with one or a few water molecules attached have
shed some light onto this topic~\cite{Korter:JPCA102:7211, Zwier:JPCA105:8827, Lippert:CPL376:40}. One of many roles of tryptophan is the radiation protection of nucleic acids, \textit{e.g.}, as a
near-UV-absorbing building block of the eumelanin polymer~\cite{Meredith:SM2:37,
   Meredith:PCR19:572}. Indole became one of the widely studied model systems to understand the
photochemistry of heteroaromatic systems, in particular, the role of the $\pi\sigma^\ast$
states~\cite{Sobolewski:PCCP4:1093}. The excited states and photodynamics of gaseous indole were
studied by a large variety of techniques, including high-resolution
spectroscopy~\cite{Berden:JCP103:9596, Brand:PCCP12:4968:2010, Kuepper:PCCP12:4980}, pump-probe
laser spectroscopy~\cite{Godfrey:PCCP17:25197, Livingstone:JCP135:194307}, ion
imaging~\cite{Lin:JCP123:124303}, and \emph{ab initio} calculations~\cite{Sobolewski:CPL315:293}.
Radiation below 300~nm has the potential to cause pyrimidine dimerization, \textit{e.g.}, leading to
melanoma~\cite{Douki:PPS12:1286}. The most damaging is radiation around 250 nm, corresponding to a
minimum of the indole absorption spectrum.

While the nucleic acids in solution dominantly suffer radiation damage indirectly, via reactions
with OH radicals and hydrated or pre-hydrated electrons formed upon irradiation of
water~\cite{Ito:IJRB63:289, Alizadeh:ARPC66:379}, they can be also ionized
directly~\cite{CRC:Lehnert:2007, Morgan:RR159:567}. Then, charge migration and transfer take place,
and indole can be one of the sinks for the positive holes formed~\cite{Milligan:NAR31:6258,
   Butchosa:OBC8:1870}. Redox properties of indole were studied for decades using kinetic
techniques. There were multiple experimental and theoretical studies of indole's photoelectron
spectrum in the gas phase~\cite{Eland:1969, Dolby:1976, Domelsmith:1977, Kubota:JESRP128:165,
   Plekan:JPCA124:4115}. Recently, the electronic structure of indole was investigated in a combined
experimental and theoretical study~\cite{Plekan:JPCA124:4115}, using tunable X-ray radiation and
\emph{ab initio} electron propagation and density functional theory. So far, the complete electronic
structure of indole in the solvent environment, and particularly in water, was not reported.
However, the resonant two-photon ionization (R2PI) spectrum~\cite{Kumar:FD212:359} of aqueous-phase indole
and the valence photoelectron spectrum of aqueous-phase tryptophan~\cite{Roy:JPCB122:3723}, with it's indole
side chain, provided information on indole's valence electronic structure.

Here, we used soft X-rays in conjunction with a liquid microjet to record the full
photoemission spectrum, including valence, core, and Auger electrons, of indole in an aqueous solution.
Previous liquid-jet photoemission studies have provided accurate insight into the
electronic structure of various molecular species, including nucleic acids, organic chromophores,
anions, cations, and transition metal complexes~\cite{Seidel:JPCL2:633, Tentscher:JPCB119:238,
   Seidel:PCCP19:32226}. Assisted by electronic structure calculation we assign the major experimental features, and we specifically disentangle non-specific bonding interactions, brought about by the long-range solvent polarization, and the specific effects, related to the granularity of the solvent environment, \textit{e.g.}, hydrogen bonding with the nearest solvent molecules. From a computational aspect, we demonstrate the wide applicability of the so-called maximum-overlap method,~\cite{Gilbert:JPCA112:13164} enabling us to model the ionized states with the standard ground-state quantum chemical methods.

\section{Experimental Methods}
\label{sec:setup}
The X-ray photoelectron-spectroscopy experiments were carried out at the P04 beamline of the
PETRA~III synchrotron-radiation facility at DESY~\cite{Viefhaus:NIMA710:151}. Experimental details
of the liquid microjet and photoelectron spectrometer as part of the ``Electronic structure of
Aqueous Solutions and Interfaces'' (EASI) setup were described elsewhere~\cite{Malerz:RSI93:015101}.
Briefly, solutions were prepared by mixing highly demineralized water
(18.2~M$\Omega\cdot\text{cm}^{-1}$) and 17~mM indole (Sigma-Aldrich, ${\mathord{>}}99$~\%, used without
further purification). Sodium chloride (50~mM) was added to minimize the streaming potential caused
by electrokinetic charging~\cite{Winter:CR106:1176, Faubel:ZPD10:269, Tang:PCCP12:2653,
   Huan:CL39:668, Preissler:JPCB117:2422, Shreve:CS4:1633, Kurahashi:JCP140:174506}.

A liquid jet of $\sim$28~$\upmu$m diameter, with a velocity of $\sim$60~m/s from a fused-silica nozzle, is generated using a high-performance liquid chromatography (HPLC) solvent delivery pump
with a constant flow rate and backing pressure. The liquid jet was captured/dumped in a vacuum using a
cryopump. The indole-water solution temperature was estimated to be in the range of 279–283~K in the
laminar part, which typically exists for 10~mm after the jet injection from the capillary
into vacuum~\cite{Winter:CR106:1176}.

The synchrotron light beam, with a photon energy of 600~eV and a focal size of 180~$\upmu$m in the
horizontal direction (parallel to the liquid jet) and 35~$\upmu$m in the vertical direction
(perpendicular to the liquid jet), intersected the jet perpendicular to the flow of the solution.
The small focal size allowed for matching the spatial overlap with the liquid jet and thereby
keeping the signal contributions from the ionization of gas-phase water molecules surrounding the jet
low. The excitation was carried out with circularly polarized light, and a backward-scattering
electron-detection geometry, corresponding to an angle of 130° with respect to the light
propagation direction, \textit{i.e.}, near magic angle, as detailed
elsewhere~\cite{Malerz:RSI93:015101}. The emitted electrons passed from the main
interaction chamber, operated at $10^{-4}$~mbar, through a 800~$\upmu$m diameter orifice to the
differentially pumped detector chamber, operated at ${\mathord{\sim}}10^{-8}$~mbar, which housed a
hemispherical electron energy analyzer equipped with a microchannel-plate detector. The small jet
diameter, in conjunction with the small distance of 800~$\upmu$m
between the liquid jet and the orifice, assured that a significant fraction of detected electrons did
not inelastically scatter with water gas-phase molecules near the jet
surface~\cite{Winter:CR106:1176, Faubel:ZPD10:269}. The energy resolution of the P04 beamline was
better than 250~meV. The energy resolution of the hemispherical analyzer was better than 200~meV.
Therefore, the total energy resolution was better than 320~meV. Tuning the detecting kinetic-energy
range of the hemispherical electron energy analyzer, we experimentally obtained the photoelectron
spectra (PE spectra) from aqueous-phase indole's valence band as well as from core ionization of the
nitrogen and carbon $1s$ orbitals, including the Auger electrons.

\section{Theoretical Methods}
\label{sec:theoretical}
Photoelectron spectra were modeled within the nuclear-ensemble method, which can be viewed as a
particular realization of the reflection principle, \textit{i.e.}, by projecting the nuclear density in the
electronic ground state onto the ionized state and further onto the spectrum of ionization
energies~\cite{Oncak:JCP133:174303, Srsen:JCTC16:6428, Srsen:2021}. The ground-state density of
indole (without water molecules) was estimated within the path-integral molecular-dynamics method, accelerated with the
quantum thermostat based on the generalized Langevin equation~\cite{Ceriotti:JCTC6:1170} (PI+GLE
method). In this way, the delocalization due to the nuclear quantum effects was taken into account.
The simulation was done at the BLYP/6-31G* level of the electronic-structure theory, representing an
efficient combination for a large number of calculations during the MD simulation. A time step of
0.5~fs was used, and four replicas were propagated. The total duration of the simulation was 23~ps,
with the first 5~ps used for equilibration. In total, 100 indole geometries were used for subsequent
calculations. PI+GLE simulations were performed using our in-house code ABIN~\cite{Slavicek:ABIN}
connected to the Terachem software (version 1.93)~\cite{Ufimtsev:JCTC5:2619, Titov:JCTC9:213}.

Core-level ionization spectra were calculated at the MP2 level of theory with the cc-pCVTZ basis set
(designed specifically for core states) for carbon and nitrogen atoms and a corresponding cc-pVTZ
basis set for hydrogen atoms. For modeling the core-ionized states, we have used the maximum-overlap
method (MOM)~\cite{Gilbert:JPCA112:13164} as implemented in the Q-Chem software (version
4.3)~\cite{Shao:MP113:184}. The aqueous solution was mimicked by the non-equilibrium polarizable-continuum model with integral equation formalism (IEF-PCM)~\cite{Cances:JCP107:3032} with the atomic
radii from universal force field (UFF)~\cite{Rappe:JACS114:10024} and the electrostatic scaling factor by which the sphere radius is multiplied $\alpha$ = 1.1. Note that IEF-PCM does not include non-electrostatic contributions and is, therefore, an approximative approach to the description of solvation. In total, core-ionization
energies from 100 sampled geometries were obtained for every atom, \textit{i.e.}, one nitrogen and eight
carbon atoms. The PE spectra of aqueous-phase indole were constructed as a sum of Gaussian functions
centered at the respective calculated value of the ionization energies for each geometry. The
standard-deviation parameter for every single Gaussian was set to 0.32~eV, obtained using the
additional broadening model~\cite{Rubesova:JCC38:427}, and corresponds to the additional broadening
arising from different configurations of solvating water molecules that were not explicitly included
in our calculations. The model connects the spectral width to reorganization energy using an approximative "universal" relaxation frequency of water after the solute's ionization. The model required the reorganization energy of water molecules solvating
indole, which we calculated as a difference between the energies calculated with the equilibrium and
non-equilibrium versions of the PCM solvation model for the optimized indole geometry. The gas-phase
PE spectra were generated with the same procedure except for the standard-deviation parameter that was set
according to Silverman's rule of thumb~\cite{Silverman:1986, Srsen:JCTC16:6428}.

The valence-ionization spectrum was calculated using the long-range-corrected Perdew--Burke--Ernzerhof
functional (LC-$\omega$PBE)~\cite{Vydrov:JCP125:234109} with the aug-cc-pVDZ basis set. The
range-separation parameter $\omega$ was optimized to the value of 0.3~$a_0^{-1}$ on the set of 100
indole geometries using the ionization-potential theorem, \textit{i.e.}, minimizing the difference between
ionization energies obtained as the energy of the HOMO and as the difference between electronic energies
of the ground and ionized states~\cite{Salzner:JCP131:231101, Muchova:JPCM31:043001}. Using the optimized
parameter $\omega$, the first two ionization energies were subsequently calculated for each geometry
with the MOM approach. The photoelectron spectrum was generated by the same procedure described above. We have tested and confirmed the robustness of our ionization energies calculations; the detailed results are presented in the SI.

Furthermore, we explored the influence of involving explicit water molecules in our calculations
on the resulting ionization energies. To do so, we (i) calculated ionization energies of solvated
clusters containing the indole molecule with one to three explicit water molecules, (ii) we executed
the molecular-dynamics simulation (distinctive from PI+GLE described above) of indole involving explicit water molecules and then calculated the
valence-ionization spectrum for snapshots of the MD simulation. Ad (i), the microhydrated clusters were
optimized at the MP2/cc-pVTZ level of theory, and core- and valence-ionization energies for the minimal
geometries were calculated using the same approach as described above. Ad (ii), the molecular-dynamics
simulation was performed for a droplet of indole in water, using the QM/MM approach with the quantum
part containing the indole molecule and the MM part consisting of 500 water molecules. The QM part was
described at the BLYP/6-31g* level with Grimme’s dispersion correction D2~\cite{Grimme:JCC27:1787},
the MM part was described with the TIP3P water model~\cite{Neria:JCP105:1902, Jorgensen:JCP79:926}.
Calculations were performed at the standard temperature of 298.15~K during the whole simulation by the Nosé–Hoover thermostat.
The QM/MM simulation was performed using our code ABIN connected to the Terachem software
(version 1.93), the initial arrangement of water molecules was obtained using the Packmol
code~\cite{Packmol:JCC30:2157}. The total duration of the simulation was 25~ps, with time steps of
0.5~fs. From the last 20~ps, 100 geometries were sampled with equidistant steps, \textit{i.e.}, every 200~fs.
We have then extracted the coordinates of the indole molecule without water molecules and with its
closest 20 water molecules, respectively. Those structures were then used to calculate the valence-ionization spectrum following the procedure described above (including PCM) to inspect the influence
of explicit water molecules on calculated valence-ionization energies.

The onset of the Auger-electron spectra was calculated as the difference
between the core-ionized and double-valence-ionized electronic energies calculated at the MP2 level
with the cc-pCVTZ basis set for N and C atoms and the cc-pVTZ basis set for H atoms. The core-ionized states were described with the MOM method. The solvation was described by the nonequilibrium PCM model.
The Auger spectra were modeled by two different approaches for a single
geometry optimized at the MP2/cc-pVTZ level. In the first approach, we evaluated the onsets of the spectra
within the MOM method with correlated wavefunctions, and then we modeled the higher transitions with
a simpler \textit{ab initio} approach based on the complete active space configuration interaction
(CAS-CI) wavefunction expansion. The Auger intensities were estimated within the qualitative
approximate scheme based on the Mulliken population analysis~\cite{Mitani:2003}. In this method,
relative transition rates are approximated using atomic populations of valence orbitals of
particular final states on core-ionized atoms. The valence orbitals are constructed by the CAS-CI
method and the cc-pVTZ-f basis on the neutral-ground-state wavefunction. The kinetic energies of the
Auger electrons $E_i$ were evaluated as
\[ E_i = E_{1s} - E_{2h,i} \,, \]
where $E_{1s}$ is the energy of the core-ionized state and $E_{2h,i}$ is the energy of the final
two-hole state. The spectrum was then shifted to match the onset described above.
The second approach for the calculation of Auger energies and intensities utilized the Feshbach--Fano approach implemented in Q-Chem 6.0 \cite{Skomorowski2021}. The initial core-ionized states are described using the fc-CVS-EOM-CCSD (frozen-core core–valence separated equation-of-motion coupled-cluster singles and doubles) method, while the final doubly ionized states are described by the EOM-DIP-CCSD (equation-of-motion double-ionization potential coupled-cluster singles and doubles) method. The uncontracted version of the aug-cc-pVDZ basis set was used, according to a recent publication \cite{Sarangi2020}. Due to the high computational cost of an EOM-DIP-CCSD method, only approximately 50 lowest-lying doubly ionized singlet and triplet final states are covered by the approach. The Auger energies and transition rates were calculated in the gas phase and then shifted by the solvent shift calculated at the same level of theory, using the PCM solvation model and MOM method for obtaining core-ionized states. In both approaches, we started with core-ionized states localized on different atoms – eight
carbons and one nitrogen in total. The line spectra consisting of energies and intensities from the contributions of final states were broadened with a Gaussian distribution with a
standard-deviation width of 2~eV to reproduce the broadening observed in the experiment.
The calculations were performed in the development-version TeraChem software (version
1.9)~\cite{Ufimtsev:JCTC5:2619, Titov:JCTC9:213} and Q-CHEM (version 6.0)~\cite{qchem5.0AndLater}.

\section{Valence Photoemission}
\begin{figure}
\centering
  \includegraphics[width=0.6\textwidth]{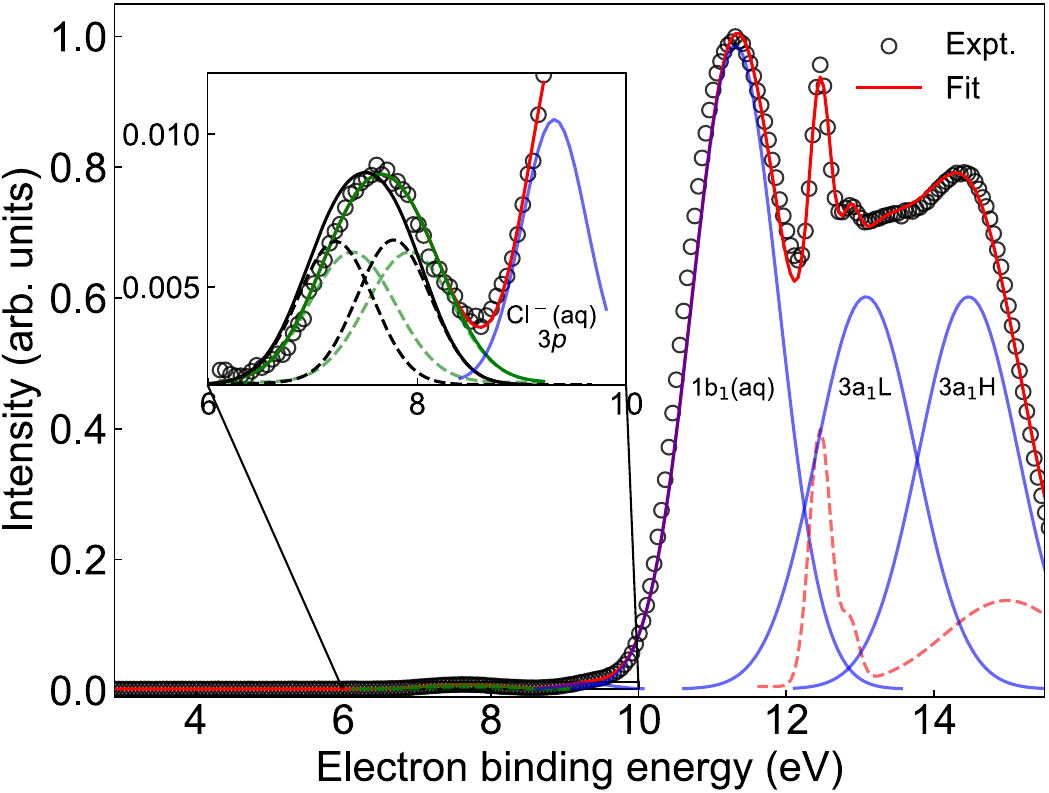}
  \caption{Experimental valence photoelectron spectra (grey empty circles) from a 17 mM indole aqueous solution containing 50 mM NaCl. The solid red line represents the overall fit. The contributions from the $1b_1$ and $3a_{1}$
      molecular orbitals of liquid water, with the $3a_{1}$ components $3a_{1}L$ and $3a_{1}H$, and
      Cl$^-$ $(3p)$ of sodium chloride background are indicated in the figure with blue solid lines.
      A dashed red line indicates the contributions from gaseous H$_2$O. The inset shows the details
      in the binding-energy region between 6 and 10~eV with the indole photoemission signal. The
      experimental data were either fitted with one Gaussian (green solid line) or with two
      Gaussians (green dashed lines) representing the signals from the HOMO and HOMO$-$1 orbitals of
      aqueous-phase indole; see text for details. The black solid line represents the calculated valence
      spectrum of aqueous-phase indole, and the black dashed lines represent the calculated components from HOMO and HOMO$-$1.}
  \label{fig:val-spec}
\end{figure}

The PE spectrum of the valence band (VB) of an indole-water solution, measured at a photon energy of
600~eV, is shown in Figure~\ref{fig:val-spec}. The raw spectrum was energy-calibrated against the liquid water $1b_1$ peak at 11.33~eV~\cite{Winter:JPCA108:2625,
   Winter:JACS127:7203, Kurahashi:JCP140:174506, Pohl:CS10:848, Gozem:JPCL11:5162,
   Perry:JPCL11:1789, Thurmer:CS12:10558}. In the present study, no low-energy cut-off measurements from an electrically biased liquid jet have been performed. This is the reason why we cannot apply the more robust method for the determination of absolute electron binding energies, as recently reported in~\cite{Thurmer:CS12:10558}. We expect an uncertainty of the reported energies to be on the order of 100–300 meV. Photoemission signals (gray empty circles) are fitted
with multiple Gaussian functions, representing the contributions from the involved atomic and
molecular orbitals, and the red solid line represents the overall fit.

By comparing the valence-band PE spectrum of the indole-water solution to that of a 50~mM NaCl
solution in water, we identified the emerging peak that exclusively corresponds to aqueous-phase indole,
shown in the inset of Figure~\ref{fig:val-spec} with fitted contributions (green lines). We used two
approaches for fitting the measurement, \textit{i.e.}, either one Gaussian or two Gaussians. The binding
energy extracted from the single-Gaussian fit is 7.65(1)~eV with a full width at half maximum (FWHM)
of 1.18~eV. The spectral width reflects, in a non-trivial way, the reorganization energy of the
solute upon single ionization. However, the observed 1.18~eV value is significantly larger than to
be expected for aromatic molecules of similar size in solution~\cite{Tentscher:JPCB119:238}.
Therefore, this immediately indicates that this band corresponds to the ionization from multiple
orbitals, \textit{i.e.}, the peak spans at least two close-lying molecular orbitals. This tentative conclusion
is supported by the corresponding gas-phase data, showing the HOMO and HOMO$-$1 orbitals to be
separated by merely 0.45~eV, see Table~\ref{tab:vie}. It is further supported by the previously reported valence photoelectron
spectrum of aqueous-phase tryptophan~\cite{Roy:JPCB122:3723} as well as by our \emph{ab initio} calculations as we
discuss below. The energies of these two orbitals (green dashed
lines in the inset of Figure~\ref{fig:val-spec} have been extracted, from a fit using two Gaussians,
to be 7.38 and 7.93~eV with a FWHM of 0.97~eV; the difference between the HOMO and HOMO$-$1 energies
was fixed to 0.55~eV based on our calculations, see details below, and the amplitudes of the two
transitions were set to be identical, assuming the same cross sections for ionization from HOMO and
HOMO$-$1 orbitals of indole. The extracted energies are in good agreement with the published photoelectron-spectroscopy data for aqueous-phase tryptophan,~\cite{Roy:JPCB122:3723} with HOMO and HOMO$-$1 energies of 7.3~eV and 8.0~eV, respectively.

\begin{table}
  \caption{Vertical ionization energies (VIE) of aqueous and gaseous indole, in units of eV.}
  \label{tab:vie}
  \centering
   \begin{threeparttable}
   \begin{tabular*}{0.6\textwidth}{l@{\extracolsep{\fill}} cccc}
    \hline
     &Exp.$_{aq}$&Calc.$_{aq}$&Exp.$_{g}$\tnote{a}&Calc.$_{g}$\\  \hline
     HOMO&7.38&7.22&7.90&7.86 \\
     HOMO$-$1&7.93&7.77&8.35&8.34 \\
    \hline
   \end{tabular*}
   \begin{tablenotes}
        \item[a] VIE reported for indole in the gas phase in ref.\cite{Plekan:JPCA124:4115}
   \end{tablenotes}
   \end{threeparttable}
\end{table}

Upon transfer of the molecule from the gas phase into an aqueous solution, the ionization energy is
expected to be shifted due to the polarizable environment~\cite{Pluharova:JPCB115:1294}. The extent
of the solvent shift is controlled by the relative stabilization of the initial neutral ground state
and the final radical-cation state in water. A typical shift observed for aromatic molecules of a
size similar to indole is on the order of 1~eV~\cite{Tentscher:JPCB119:238}.

Our calculated valence-ionization energy in the gas phase agrees excellently with the experimental
data of~\citet{Plekan:JPCA124:4115} with a discrepancy in binding energies of less than 0.05~eV, see
Table~\ref{tab:vie}. As for the calculated aqueous-phase valence spectrum including HOMO and HOMO$-$1
contributions, Figure~\ref{fig:val-spec} (black solid line in the inset), has its maximum at 7.50~eV,
differing from our experimental aqueous-phase data by 0.15~eV. This is a very good agreement; the
remaining discrepancy is to be attributed to the treatment of solvation. The observed shift in the
present case is on average $\sim$0.47~eV for the HOMO and HOMO$-$1 orbitals; see Table~\ref{tab:vie}.
The exact value slightly depends on the fitting procedure. However, it is surprisingly small,
suggesting competing specific and non-specific solvent effects. The calculated spectrum exhibits the
spectral shape, which is in good agreement with the present experiment. Note that the calculated
spectra were generated within the cluster-continuum model, \textit{i.e.}, the solvent was described as a
polarizable continuum while, simultaneously, the 20 nearest water molecules were included in the
calculations. If these explicit molecules were excluded, the maximum of the calculated photoelectron
peak would be shifted to 7.38~eV, compared to 7.50~eV. This identified certain specific solvent
effects, manifested by a difference of 0.12~eV. These specific solvent effects counter, to some extent, the non-specific effects and are responsible for the relatively small
solvent shift observed in the experiment compared to similar molecules.

\begin{figure}
\centering
   \includegraphics[width=0.6\textwidth]{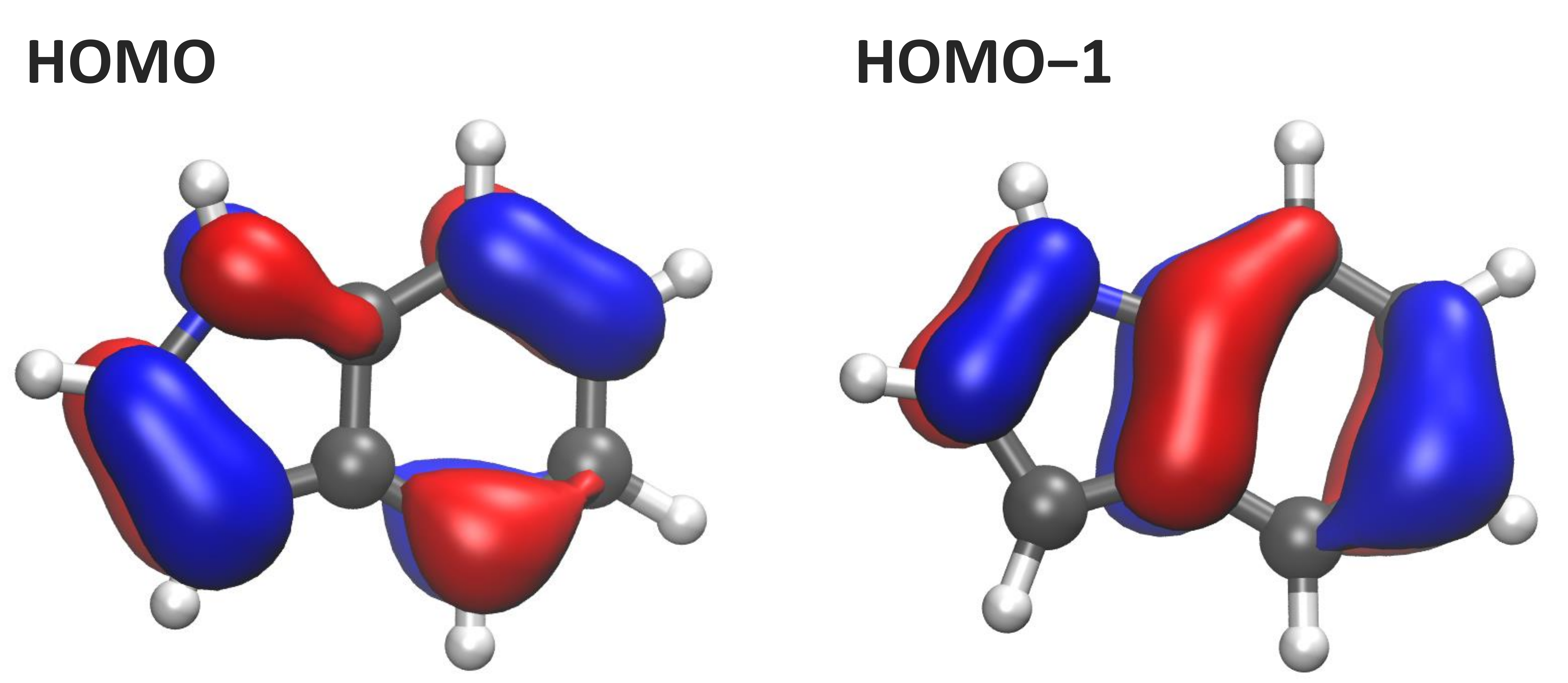}
   \caption{Calculated highest occupied molecular orbital (HOMO) and second-highest occupied
      molecular orbital (HOMO$-$1) of aqueous-phase indole. Both orbitals are of $\pi$ character and are
      delocalized over the molecule.}
   \label{fig:orbitals}
\end{figure}

The calculations reveal that the aqueous-phase valence band actually consists of two transitions
corresponding to two different orbitals, both of them of $\pi$ character, see
Figure~\ref{fig:orbitals}. The difference in the peak positions based on our calculations is estimated
to be 0.55~eV. This is somewhat larger than the 0.45~eV splitting for the gas-phase experiment,
which is reproduced very well by our \emph{ab initio} calculations. However, the discrepancy is not
significant when we consider the inaccuracies in the calculations and the errors introduced by the
fitting procedure for the experimental data.

\begin{figure}
\centering
   \includegraphics[width=0.45\textwidth]{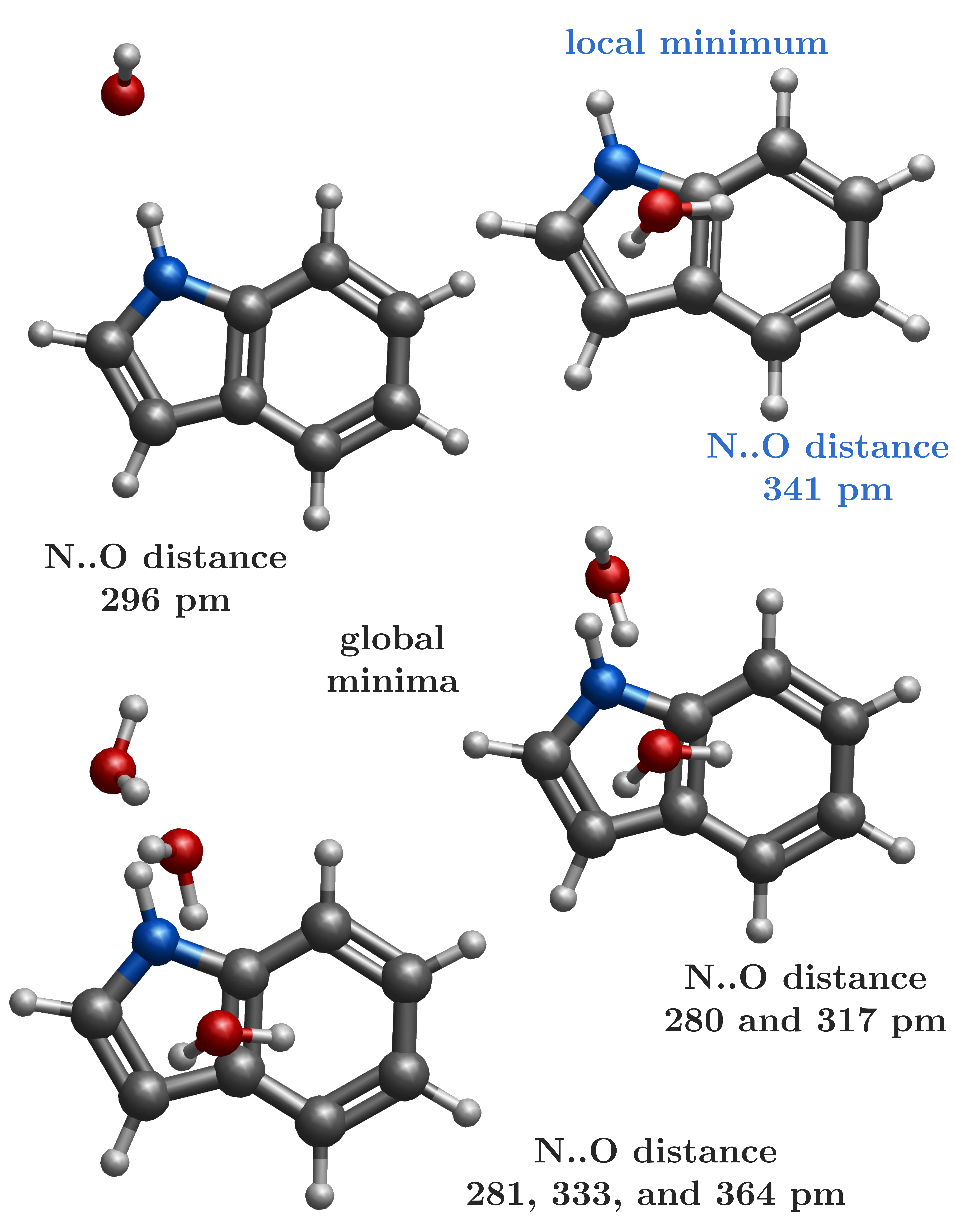}
   \caption{Optimized structures of gas-phase indole-water clusters containing one, two, and three
      water molecules. At least one water molecule forms a hydrogen bond with the N–H group while
      others are pointing towards the $\pi$ ring.}
  \label{fig:indole-water-structure}
\end{figure}

Let us now focus on the interplay between specific and non-specific solvent interactions. An
interesting insight is brought about by the inspection of the indole-water dimer complex in the gas
phase. The global minimum of this complex corresponds to the hydrogen-bonded structure via the
$\text{N–H}\cdot\text{O}$ contact~\cite{Korter:JPCA102:7211}, see also
Figure~\ref{fig:indole-water-structure}. This complex has a calculated vertical ionization energy (VIE)
of 7.51~eV and an adiabatic ionization energy (AIE) of 7.27~eV. These calculations can be compared
with the available experimental data, where R2PI experiments provided AIEs of the bare indole
molecule as well as the cluster of indole with one water molecule of 7.76 and 7.34~eV,
respectively~\cite{Hager:CPL113:503}. The energetic drop of about 0.4~eV in experimental AIE when
adding the first water molecule is also observed in our calculations of VIEs, which drop from 7.87
to 7.51 eV, see 
Table~\ref{tab:IE}. 
Water molecules can, however, also bind to the $\pi$ system of the
indole molecule; see Figure~\ref{fig:indole-water-structure}. The hydrogen-bonded complex is
energetically preferred by 0.1~eV, according to our calculations. The VIE for this complex is calculated
to be 0.6~eV larger while the adiabatic energy is only 0.3~eV larger than for the $\pi$-system-bound
structure. Thus, the reorganization energy is larger for the dimer with the water molecule bound to
the $\pi$ system of indole than for the hydrogen-bonded system. At the same time, the optimized
structure of the oxidized indole-water complex is very different from the neutral structure in this
case.

In liquid water, the water molecules are delocalized over the whole indole molecule. One of the
water units is certainly located in the hydrogen-bonded position as it is documented by the
core-level ionization spectrum; see section \emph{``Core-electron binding energies``} below. Most
other water molecules are, however, placed above or below the $\pi$ ring of the indole system. In
those positions, the water molecules will need to dramatically reorganize upon the ionization, which
will lead to pronounced reorganization energy and decreasing value of the Franck--Condon overlap
for the states energetically close to the AIE. The water molecules surrounding the $\pi$ system also
contribute to the increase of the VIE, counterbalancing the VIE decrease by long-range polarization. As
a result, the solvent shift is smaller than expected from the dielectric theory or from our
experience with organic molecules of a similar size.

\subsection{Core-Level Binding Energies}
\label{sec:core-BE}

\begin{figure}
\centering
   \includegraphics[width=0.6\textwidth]{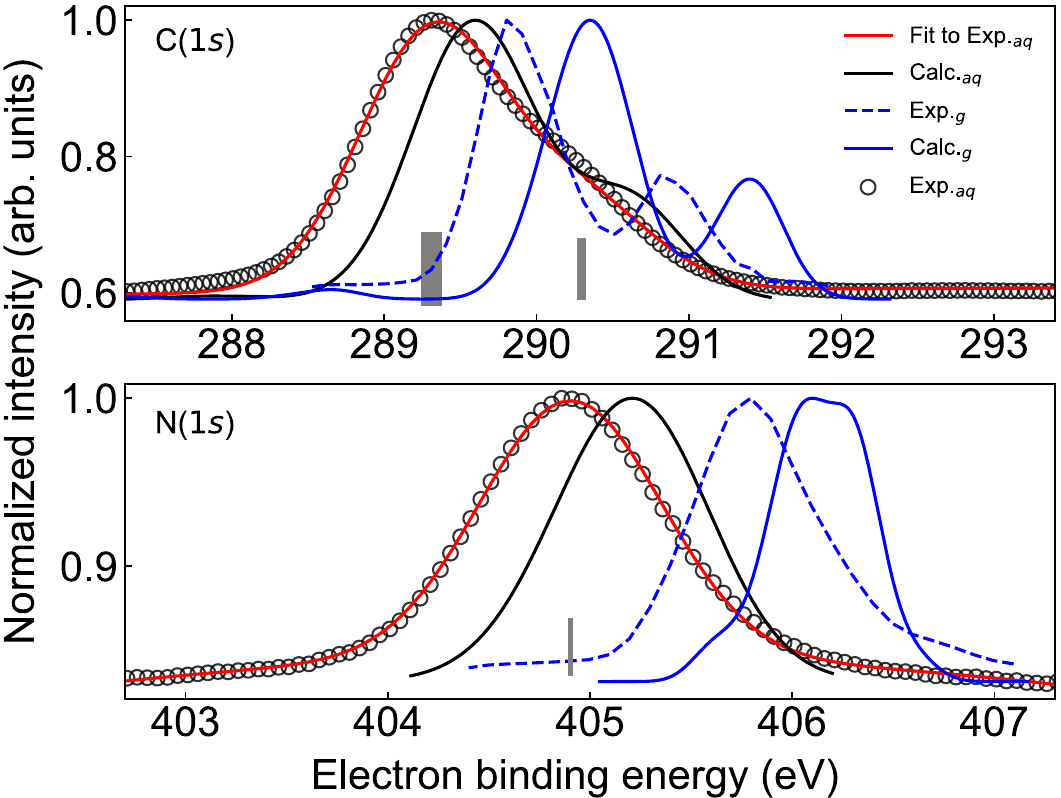}
   \caption{C($1s$) (upper panel) and N($1s$) (lower panel) photoelectron spectra of indole. Grey
      empty circles represent the experimental data and red solid lines represent the overall fit
      for aqueous-phase indole. The light-gray bars indicate the central positions of the individual
      Gaussians used to fit. The black and blue solid lines represent the simulated spectra for
      aqueous-phase indole and gas-phase indole, respectively. The blue dashed lines are experimental data
      reported for indole in the gas phase.\cite{Plekan:JPCA124:4115}}
   \label{fig:core-spec}
\end{figure}

The carbon and nitrogen $1s$ core-level PE spectra of aqueous-phase indole, measured at 600 eV photon energy, are shown in Figure~\ref{fig:core-spec} together with a fit of a sum of multiple Gaussians. Spectra are displayed on an electron binding energy (BE) axis, and the raw spectra were background-corrected by subtracting second-order polynomials to account for the contributions of electrons scattered in the solution. As in the case of the valence spectra, BE calibration was performed against the liquid water 1\emph{b}$_1$ (HOMO) peak at 11.33 eV, and we expect a 100–300 meV uncertainty of the energies reported here.

The C($1s$) PE spectrum of gaseous indole showed two distinct features, one centered at 289.89~eV arising
from the six carbon atoms not bound to nitrogen and one centered at 290.86~eV arising from the two
carbon atoms directly bound to nitrogen~\cite{Plekan:JPCA124:4115}, based on gas-phase results. For
aqueous-phase indole, these two features largely merge, and we fitted the spectrum simultaneously
with two groups of Gaussian functions ($6+2$). Thus, the characteristic fitted parameters were mainly
the two independent peak positions. The FWHM for each Gaussian function was fit to be 1.12~eV, which
is fully consistent with values typically found for the photoionization of neutral molecules of a
similar size in solution. For the N($1s$) PE spectrum, resulting from the single nitrogen atom in indole,
one Gaussian centered at 404.90~eV with a FWHM of 1.06~eV was used to fit the spectrum. Compared to
the reported gas-phase data~\cite{Plekan:JPCA124:4115}, we obtained solvent-induced shifts of
$\sim$0.57~eV for C($1s$) and $\sim$0.92~eV for N($1s$).

The N($1s$) and C($1s$) BEs for aqueous-phase and gaseous indole from both the experiment and our
calculations are shown in Table~\ref{tab:BE}. Note that the calculated values for C($1s$) were
obtained by fitting the simulated C($1s$) core-level PE spectrum with the same fitting procedure as
for the experimental spectrum. The FWHM of each Gaussian in the fit for the C($1s$) simulated
spectrum is 0.92~eV. The simulated N($1s$) spectrum for aqueous-phase indole is centered at 405.28~eV with
a FWHM of 0.58~eV. The calculated PE spectra for the N($1s$) and C($1s$) electrons from gas-phase and
aqueous-phase indole were shifted with respect to the experiment by $\sim$0.3--0.4~eV,
which is quite acceptable for these large absolute energies. More importantly, the calculations well
reproduce the shapes of the spectra in both the gas and liquid phase, including the broadening upon
solvation and the solvent shift; see Figure~\ref{fig:core-spec}. Even the C($1s$) spectrum
corresponding to convoluted signals from eight different carbon-atom sites in the molecule is
described well for both aqueous and gas-phase indole.

\begin{table}
  \caption{C($1s$) and N($1s$) experimental and calculated binding energies (eV) for aqueous and gas-phase indole.}
  \label{tab:BE}
  \centering
  \begin{threeparttable}
  \begin{tabular*}{0.6\textwidth}{l@{\extracolsep{\fill}} cccc}
    \hline
     Peak & Exp.$_\text{aq}$ & Calc.$_\text{aq}$ &Exp.$_\text{g}$\tnote{a}&Calc.$_\text{g}$\\  \hline
     {\quad}C$^{1-6}$&289.31&289.59&289.89&290.34\\
     {\quad}C$^{7-8}$&290.29&290.58&290.86&291.39\\
     {\quad}N$^1$&404.90&405.28&405.82&406.15\\
    \hline
  \end{tabular*}
  \begin{tablenotes}
        \item[a] Gas-phase ionization energies reported in ref.\cite{Plekan:JPCA124:4115}
  \end{tablenotes}
  \end{threeparttable}
\end{table}

\subsection{Solvation Structure}
While valence electrons are to a large extent delocalized, core electrons are localized near one of
the nuclei. X-ray PES is thus suited for investigating the local or nearest-neighbor hydrogen-bonding
structures by analyzing core-level chemical and solvent shifts as well as peak profiles arising from the
intermolecular interaction of indole with surrounding water molecules~\cite{Winter:NIMA601:139}.

The solvent-induced shifts for the valence, C($1s$), and N($1s$) signals are significantly different. This is
indicative of the specific solvation structure near the ionized molecules. Energy shifts are approximately
0.47~eV for the valence ionization, 0.57~eV for the C($1s$) ionization, and 0.92~eV for the N($1s$) ionization.

When we examine the geometries of the microhydrated indole, Figure~\ref{fig:indole-water-structure}, it
is clear that one closest water molecule tends to approach the nitrogen atom to form a direct
$\text{N-H}\cdots\text{O}$ hydrogen bond, reflecting the structure of the gas-phase indole-water
dimer~\cite{Korter:JPCA102:7211}. This leads to a larger solvent stabilization of the core-ionized
nitrogen atom as the water molecule's dipole is pointing at indole's N, \textit{i.e.}, the water is oriented
toward N-H with its oxygen atom. On the other hand, core-ionized carbons are destabilized by the
arrangement of the other water molecules. Those waters are hydrogen-oriented toward the indole ring.
Therefore, the positively charged core holes on carbon atoms created by the ionization interact with
partial positive charges of water’s hydrogen atoms.

This interpretation is further supported by calculations showing how calculated core- and
valence-ionization energies of indole change upon adding explicit solvent molecules and introducing
the polarizable continuum representing bulk water, see Table~\ref{tab:IE}. We observed that the
addition of a single water molecule coordinated to the nitrogen atom causes a large decrease in the
N($1s$) binding energy, and this holds even when placing the system into a dielectric environment.
The effect of this hydrogen-bonded water molecule on the valence- and C($1s$)-electron energies is
much smaller. This is related to the delocalized character of the valence-electron hole and to the
diffuse hydrogen-bond arrangements around the core-ionized carbon atoms, respectively.

\begin{table*}
\footnotesize
   \caption{Calculated ionization energies (eV) of gas-phase and aqueous-phase indole containing
      0--3 explicit water molecules. 'Solvated cluster' refers to a system placed into a
      dielectric environment to mimic non-specific solvation effects. For clarity, C$(1s)$
      ionization energies are averaged for six (1--6) and two (7--8) carbon atoms, which exhibit
      very similar energies.}
   \label{tab:IE}
   \centering
   \tabcolsep=0.09cm
   \begin{tabular}{cccccccccccccc}
     \hline
     &  &\multicolumn{5}{c}{\textbf{Gas-phase cluster}}  &  &  & \multicolumn{5}{c}{\textbf{Solvated cluster}}  \\
     \cmidrule(l{8mm}r{8mm}){3-7} \cmidrule(l{8mm}r{8mm}){10-14}
     \textit{n}$_{water}$  &  & HOMO & HOMO$-$1 & N($1s$) & C$^{1-6}$($1s$) & C$^{7-8}$($1s$) &  &  & HOMO & HOMO$-$1 & N($1s$) & C$^{1-6}$($1s$) & C$^{7-8}$($1s$) \\
     \cmidrule(l{3mm}r{2mm}){3-7} \cmidrule(l{3mm}r{2mm}){10-14}
     0 &  & 7.87 & 8.29 & 406.24 & 290.15 & 291.24 &  &  & 7.13 & 7.57 & 405.28 & 289.39 & 290.43  \\
     1 &  & 7.51 & 7.94 & 405.55 & 289.82 & 290.82 &  &  & 7.03 & 7.48 & 404.95 & 289.32 & 290.30 \\
     2 &  & 8.02 & 8.42 & 405.99 & 289.72 & 291.28 &  &  & 7.22 & 7.64 & 405.15 & 289.46 & 290.45 \\
     3 &  & 7.93 & 8.33 & 405.74 & 290.21 & 291.17 &  &  & 7.18 & 7.60 & 404.98 & 289.43 & 290.40 \\
     \hline
   \end{tabular}
\end{table*}

\subsection{Auger-Electron Spectra}

\begin{figure}
\centering
   \includegraphics[width=0.6\textwidth]{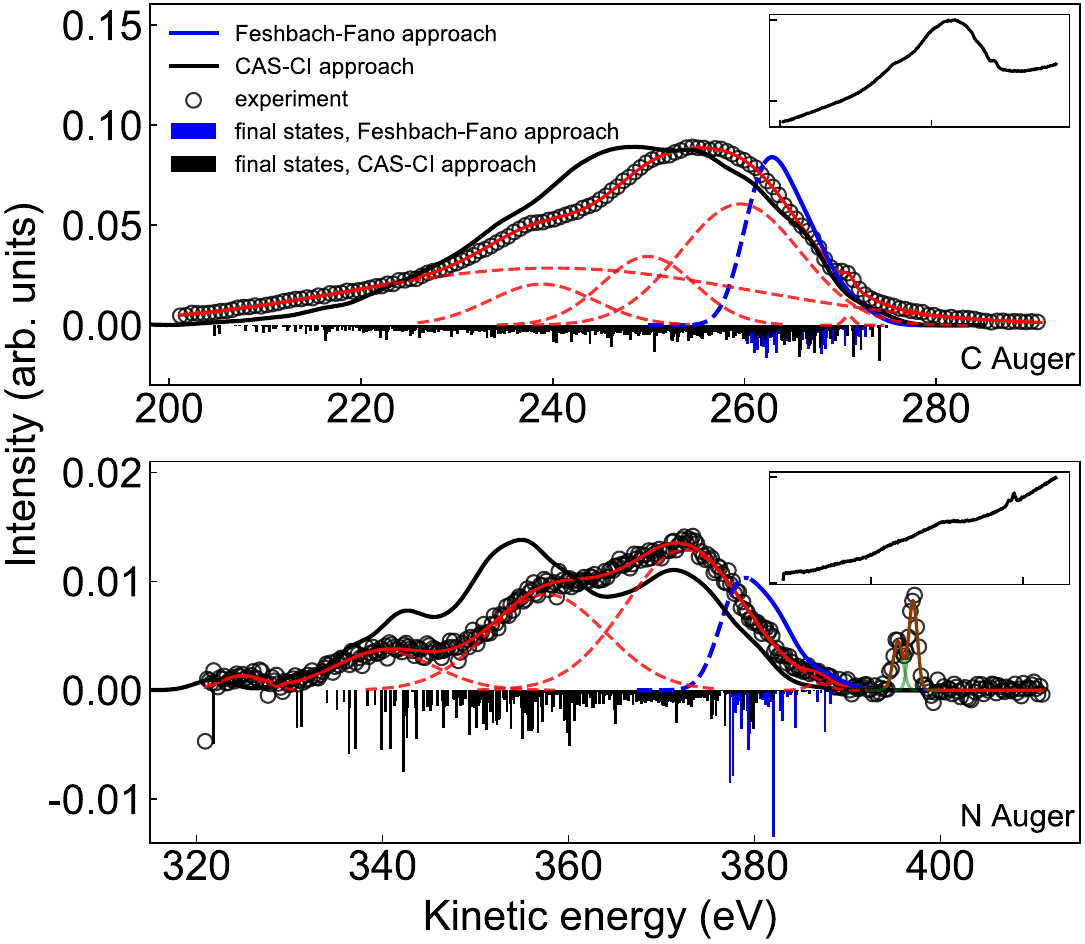}
   \caption{C-Auger and N-Auger spectra (black circles) of aqueous-phase indole after fitted background
      subtraction. Red solid lines represent the overall fit and the
      red dashed lines are the individual fitted Gaussians. The black solid lines represent the simulated
      carbon/nitrogen $1s$ Auger spectra via the CAS-CI approach, the blue solid lines represent simulated carbon/nitrogen Auger spectra via the Feshbach-Fano approach. Dashed blue lines describe a part of modeled spectra, for which a significant contribution of higher-lying doubly ionized states is expected, but is absent due to the method limitations. Contributions from the ionization of Cl$(2p)$ are included in the fit
      and shown as green solid lines. The downward-facing sticks schematically represent the population of involved two-hole states. The insets show the raw experimental Auger
      spectra. A global fit using second-order polynomials for the contributions of the scattered-electron
      background, with a highly asymmetric spectral structure, was applied.}
   \label{fig:Auger}
\end{figure}

Measured Auger-electron spectra following N($1s$) and C($1s$) core-level ionization of aqueous-phase indole at 600 eV photon energy are shown in Figure~\ref{fig:Auger} top and bottom (open circles), respectively. Here, we have subtracted a linear background from the as-measured spectra shown in the respective insets. Both spectra are broad, approximately 60–70 eV wide. Similar to that for gaseous indole\cite{Plekan:JPCA124:4115}, the broadening includes the finite line width of the X-ray PE spectrum, the vibrational distributions of the core-hole ground and final states, the core-hole lifetime, the analyzer energy resolution, and the contributions from the solvation for aqueous-phase indole due to the conformational distribution of solvating molecules. The spectra were fitted with five Gaussians for both C-Auger and N-Auger data, corresponding to a minimal number of functions yielding a reasonable total fit of the experiment. Note though that these Gaussians have no physical meaning, and only the high-KE part of the spectrum will be analyzed here. The applied photon energy is well above the respective core-level energies such that participator and spectator Auger channels do not need to be considered. The sharp doublet peak near 396 eV kinetic energy (KE), next to the N-Auger spectrum, arises from Cl$^-(2p)$ core-level ionization; as explained in the Methods section, 50 mM NaCl has been added to the indole aqueous solution to counteract electrokinetic charging.

The high-kinetic-energy onsets of N($1s$) and C($1s$) Auger-electron spectra are roughly 388 eV and 274 eV, respectively, in accordance with the calculated values as shown in Table~\ref{tab:Auger-BE}. The KE was calibrated with reference to the energies of the C($1s$) and N($1s$) photoelectron peaks, as explained above. The measured Auger spectra have been interpreted with the help of \emph{ab initio} modeling, using the approach based on Mulliken population analysis. In addition, we used the more sophisticated yet more costly method based on the Feshbach-Fano approach. In either case, Auger peaks are found to be formed by transitions involving a large number of final states, and the peaks can thus not be interpreted as a result of a single decay channel. The total calculated Auger-electron spectra using Mulliken population analysis, shown as solid black lines in Figure~\ref{fig:Auger}, are seen to reasonably well reproduce both experimental spectra, even exhibiting the experimentally observed substructure. The solid blue lines in Figure~\ref{fig:Auger} show the calculated Auger spectra using the Feshbach-Fano approach. This allowed us to directly estimate both nitrogen and carbon leading Auger-electron contributions, including the respective high-energy edges. However, for a molecule of the size of indole (and larger), only a limited number of final doubly ionized states can be captured due to the computational cost, not allowing to model the whole Auger spectra. 
This is reflected in the dashed blue curve, below which spectra can no longer be reliably calculated due to the missing calculated final states. We can conclude that a much simpler approach based on Mulliken population analysis reproduces the whole spectrum, and this method is sufficient for the present study.

\begin{table}
  \caption{Summary of the measured and calculated BEs (eV) for valence, N($1s$), and C($1s$) ionization, and
      the KE onsets of the Auger electrons, for aqueous-phase indole.}
  \label{tab:Auger-BE}
  \begin{tabular*}{0.7\textwidth}{l@{\extracolsep{\fill}} cccc}
    \hline
     &\multicolumn{2}{c}{\textbf{Exp. fitting}}  & \multicolumn{2}{c}{\textbf{Simulations}}  \\
     \cmidrule(l{3mm}r{1mm}){2-3} \cmidrule(l{3mm}r{1mm}){4-5}
     & BE & KE onset & BE  & KE onset \\  \hline
     HOMO            & 7.38   &--      & 7.22    & --      \\
     HOMO$-$1          & 7.93   &--      & 7.77    &--       \\
     N($1s$)          & 404.90 &--      & 405.28  &--       \\
     C$^{1-6}$($1s$)  & 289.31 &--      & 289.59  &--       \\
     C$^{7-8}$($1s$)  & 290.29 &--      & 290.58  &--       \\
     N Auger         & --     &387.53  &--       & 387.02  \\
     C Auger         & --     &274.31  &--       & 272.26  \\
    \hline
  \end{tabular*}
\end{table}

The electron holes of the final state correspond to the ejected $\pi$ electrons of the indole ring; see HOMO and HOMO$-$1 orbitals in Figure~\ref{fig:orbitals}. Calculated values for gas-phase indole are 384.74 eV and 269.79 eV for N- and C-Auger electrons, respectively, in good agreement with recently published experimental data~\cite{Plekan:JPCA124:4115}. The calculated solvent shift for nitrogen and carbon Auger energies is 2.28 and 2.47 eV, respectively, confirming the higher N($1s$) core-ionized solvent stabilization relative to the C($1s$) core-ionized state.

In conclusion, we have provided the first full photoemission spectrum of indole in aqueous solution by measuring the
valence and core-level photoelectron and Auger spectra following ionization by 600~eV synchrotron
radiation. Experimental spectra are interpreted with the help of high-level \emph{ab initio}
calculations. All characteristic peaks of aqueous-phase indole were assigned and the explicit and
global solvent-induced energy shifts were extracted and supported by the calculations.

The lowest-binding-energy valence photoelectron peaks correspond to the ionization of the HOMO and
{HOMO$-$1} electrons, with binding energies of 7.38 and 7.93~eV, respectively. The observed
solvent-induced shifts were relatively small in comparison with other neutral molecules of a similar
size. This is due to delocalized valence electrons and competing specific and non-specific solvation
effects. 

Indole, in the gas phase, is known to form strong hydrogen-bonded complexes with water with its N-H
group serving as the hydrogen-bond donor~\cite{Korter:JPCA102:7211}. Our results demonstrate that
this motif is also dominant in aqueous solution. Specifically, disentangling the solvent-induced
shifts, which are specific extensions of the general chemical shift, in the core-ionization spectra
enabled us to elucidate the solvent structure around the indole molecule. The core-level binding
energies for nitrogen and carbon $1s$ electrons clearly indicated the presence of specific solvent
effects due to a strong hydrogen bonding to nitrogen together with further non-specific effects due
to solvent polarization. On the one hand, there is a strong directed and specific hydrogen bonding
$\text{N-H}\cdots\text{OH}_2$, while on the other hand, there are unstructured interactions of the
water solvent with the overall molecular structure.

Furthermore, we reported and interpreted the Auger spectra, which exhibit larger solvent shifts than
the direct photoelectrons. These Auger-electron signals are brought about by many final dicationic
states. A computational technique based on electron-population analysis was demonstrated as an
efficient tool for modeling Auger spectra, aiding in further analysis of molecules in complex
environments using X-ray photoemission spectroscopy.

Overall, our detailed investigation of the photoemission spectrum of indole in water provided a
clear and refined view on the solvation of indole, specifically for its different moieties, including
a surprising highly specific single-solvent molecule-binding motif combined with further unspecific
solvent interactions. From the perspective of computational chemistry, our work demonstrated the
wide applicability of the maximum-overlap method, enabling us to model the ionic states through
standard ground-state quantum chemical methods, combined with the non-equilibrium dielectric
modeling of the environment.

\section*{Associated Content}
\subsection{Data and Code Availability}
The data of relevance to this study have been deposited on Zenodo at
\href{https://dx.doi.org/10.5281/zenodo.6519525}{DOI:~10.5281/zenodo.6519525}.

Code availability: The ABIN code for molecular dynamics v.~1.1-alpha was used, available at
\url{https://github.com/PHOTOX/ABIN}. The Packmol code v.~18.169 was used, available at
\url{http://leandro.iqm.unicamp.br/m3g/packmol/download.shtml}. Terachem v.~1.93 was used, available
at \url{https://store.petachem.com}. Q-Chem v.~4.3 and 6.0 was used, available at
\url{https://www.q-chem.com}.

\subsection{Supporting Information}
Calculated gas-phase photoelectron spectrum of indole, details on modeling Auger spectra, and robustness test of ionization energy calculations.

\section*{Author Information}
\subsection{Corresponding Authors}
\noindent \textbf{P. Slav\'{i}\v{c}ek} -- Department of Physical Chemistry, University of Chemistry and Technology, Technick\'{a} 5, 16628 Prague, Czech Republic; \href{https://orcid.org/0000-0002-5358-5538}{orcid.org/0000-0002-5358-5538}; Email: Petr.Slavicek@vscht.cz

\noindent \textbf{B. Winter} -- Molecular Physics, Fritz-Haber-Institut der Max-Planck-Gesellschaft, Faradayweg 4-6, 14195 Berlin, Germany; \href{https://orcid.org/0000-0002-5597-8888}{orcid.org/0000-0002-5597-8888}; Email: winter@fhi-berlin.mpg.de

\noindent \textbf{J. Küpper} -- Center for Free-Electron Laser Science, Deutsches Elektronen-Synchrotron DESY, Notkestraße~85, 22607 Hamburg, Germany; Center for Ultrafast Imaging, Universität Hamburg, Luruper Chaussee 149, 22761~Hamburg, Germany; Department of Physics, Universität Hamburg, Luruper Chaussee 149, 22761 Hamburg, Germany; \href{https://orcid.org/0000-0003-4395-9345}{orcid.org/0000-0003-4395-9345}; Email: winter@fhi-berlin.mpg.de

\subsection{Authors}
\noindent \textbf{L. He} -- Center for Free-Electron Laser Science, Deutsches Elektronen-Synchrotron DESY, Notkestraße~85, 22607 Hamburg, Germany; Institute of Atomic and Molecular Physics, Jilin University, 130012 Changchun, China; \href{https://orcid.org/0000-0002-6577-8716}{orcid.org/0000-0002-6577-8716}

\noindent \textbf{L. Toman\'{i}k} -- Department of Physical Chemistry, University of Chemistry and Technology, Technick\'{a} 5, 16628 Prague, Czech Republic; \href{https://orcid.org/0000-0003-2547-2488}{orcid.org/0000-0003-2547-2488}

\noindent \textbf{S. Malerz} -- Molecular Physics, Fritz-Haber-Institut der Max-Planck-Gesellschaft, Faradayweg 4-6, 14195 Berlin, Germany; \href{https://orcid.org/0000-0001-9570-3494}{orcid.org/0000-0001-9570-3494}

\noindent \textbf{F. Trinter} -- Molecular Physics, Fritz-Haber-Institut der Max-Planck-Gesellschaft, Faradayweg 4-6, 14195 Berlin, Germany; Institut für Kernphysik, Goethe-Universität Frankfurt, Max-von-Laue-Straße 1, 60438 Frankfurt am Main, Germany; \href{https://orcid.org/0000-0002-0891-9180}{orcid.org/0000-0002-0891-9180}

\noindent \textbf{S. Trippel} -- Center for Free-Electron Laser Science, Deutsches Elektronen-Synchrotron DESY, Notkestraße~85, 22607 Hamburg, Germany; Center for Ultrafast Imaging, Universität Hamburg, Luruper Chaussee 149, 22761~Hamburg, Germany; \href{https://orcid.org/0000-0002-1895-3868}{orcid.org/0000-0002-1895-3868}

\noindent \textbf{M. Belina} -- Department of Physical Chemistry, University of Chemistry and Technology, Technick\'{a} 5, 16628 Prague, Czech Republic; \href{https://orcid.org/0000-0002-0242-6879}{orcid.org/0000-0002-0242-6879}

\noindent \textbf{Notes}

\noindent The authors declare no competing financial interest.

\section*{Acknowledgements}
We acknowledge Claudia Kolbeck for the help during the experimental campaign. We thank the
PETRA~III P04 beamline staff and Moritz Hoesch in particular as well as the DESY photon science
chemistry-laboratory and crane operators for their assistance.

We acknowledge support by Deutsches Elektronen-Synchrotron DESY, a member of the Helmholtz
Association (HGF), and the use of the Maxwell computational resources operated at Deutsches
Elektronen-Synchrotron DESY. Parts of this research were carried out at PETRA~III at DESY; beamtime
was allocated for proposal II-20180012. This work was supported by the Cluster of Excellence
``Advanced Imaging of Matter'' (AIM, EXC~2056, ID~390715994) of the Deutsche Forschungsgemeinschaft
(DFG) and by the European Research Council under the European Union's Seventh Framework Program
(FP7/2007-2013) through the Consolidator Grant COMOTION (614507). L.H.\ acknowledges support by the
National Natural Science Foundation of China (Grants No.92261201, No.11704147) and a fellowship within the framework of the
Helmholtz-OCPC postdoctoral exchange program. P.S., L.T., and M.B.\ thank the Czech Science
Foundation (EXPRO project no.~21-26601X). S.M., F.T., and B.W.\ acknowledge funding from the European Research Council (ERC) under the European Union’s Horizon 2020 research and innovation program under Grant Agreement No. GAP 883759–AQUACHIRAL. F.T.\ and B.W.\ acknowledge support by the MaxWater initiative of the
Max-Planck-Gesellschaft.

\bibliography{bibliography}

\end{document}